# Low-cost automated spin coater and thermal annealer for additive prototyping of multilayer Bragg reflectors

Nathan J. Dawson[*,1,2], Yunli Lu[1], Zoe Lowther[1], Jacob Abell[1], Nicholas D. Christianson[1], Aaron W. Weiser[3], and Gioia Aquino[4]

[1]College of Natural and Computational Sciences, Hawaii Pacific University, Kaneohe, HI 96744, USA
[2]Department of Physics and Astronomy, Washington State University, Pullman, WA 99164, USA
[3]Department of Physics and Astronomy, Youngstown State University, Youngstown, OH 44555, USA
[4]College of Engineering, University of Hawaii, Honolulu, HI 96822, USA

## Abstract

We present and implement a design for an automated system that fabricates multilayer photonic crystal structures. The device is constructed with low-cost materials. A polystyrene/cellulose acetate multilayer Bragg reflector was fabricated to confirm the device's capability. A distributed feedback laser was also fabricated and characterized. The system has also been used to fabricate microlasers for a Modern Physics laboratory assignment in which students measure fluorescence, amplified spontaneous emission, lasing from one-dimensional Bragg reflectors, and lasing from scattering media.

## 1 Introduction

Photonic crystals are an important class of materials because they interact strongly with visible light through wave interference. [1, 2] They are observed in nature [3–7] and also find many uses in science and technology. [8–10] One-dimensional photonic crystal are used for antireflection coatings, [10] vapor sensors, [9, 11] and other photonic devices such as distributed feedback lasers. [8, 12–14] Inorganic and hybrid materials are often used to fabricate multilayer photonic crystals, [15–17] but polymers offer an inexpensive alternative to many costly materials and their morphologies are more easily manipulated at the nanoscale. Thus, polymers have emerged as a cost-effective option for fabricating photonic materials. [18]

Spin coating and annealing each individual layer "by hand" can be a serious undertaking where each prototype film can require hours to days of repetitive steps. Thus, automating the process can greatly decrease the burden on a student or researcher. In this paper, we describe a low-cost, turnkey system to spin/anneal polymer multilayer films that can be created from common consumer electronics parts and materials found at local hardware stores. Two applications of multilayer photonic crystals fabricated with the automated system are also presented – a visibly reflective Bragg mirror followed by a distributed feedback (DFB) laser. Implementation of the DFB lasers into a coherent light emission assignment for a modern physics laboratory course is also discussed.

---

[*]ndawson@hpu.edu

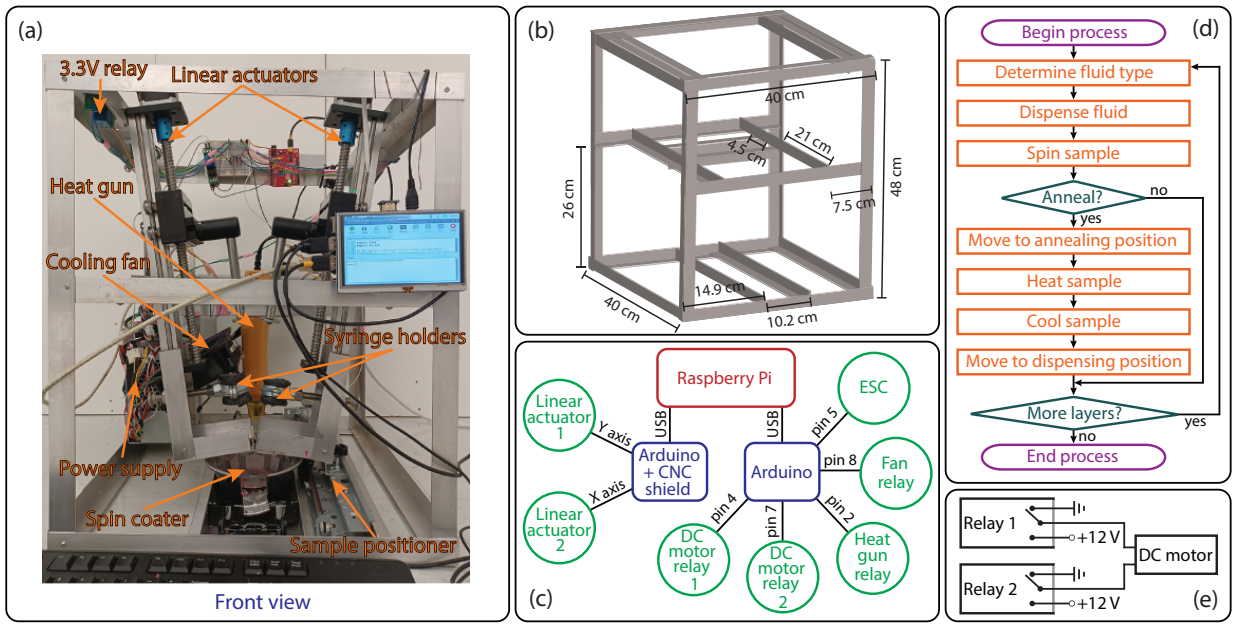

Figure 1: (Color online) (a) An image of the assembled multilayer spin coater/annealer system and (b) the frame constructed with aluminum angles. (c) Diagram showing the electronic component connections and (d) flow chart. (e) The solid-state relay circuit controlling the DC motor used to relocate the spin coater between dispensing and annealing positions.

## 2 Device Design

We describe an automated spin coater and annealer from materials accessible to most educators and researchers below.

### 2.1 Body/Frame

The outer box of the frame is shown in Figure 1(a)(b), where images of the constructed device from different perspectives are shown in Section S1 of the supplementary sections. The frame was constructed of 1/8" thick aluminum angles with 1" legs. Construction angles are manufactured using a standard set of imperial dimensions, and they are sold at local hardware stores all over the United States. Construction angles can be made from a variety of materials and we chose aluminum due to the ease of cutting and drilling relative to a harder material such as steel. Aluminum angles are usually sold with standard thicknesses of 1/16" increments. The 1/16" thick aluminum angles can be quite flimsy and the authors recommend using a minimum angle thickness of 1/8" when using aluminum as the material. Note that the cost per foot of angles increases with the thickness so finding the minimum thickness necessary for your device can save on the final cost of your device. Also note that the 1/8" thickness allowed for some bending away from a 90° angle with a reasonable amount of torque when mounting the linear actuators at a shallow angle with respect to the vertical direction. Thus, by choosing a 1/8" thickness, we were able to use the aluminum angles for both the outer frame and the structures used to support actuating components. Construction angles can also be sold in a variety of leg sizes, where 1" legs at 1/8" thickness allowed for rigid frame construction and component mounting without changing sizes; purchasing 8' standard lengths and cutting the dimensions shown in Figure 1(b) resulted in residual pieces that were used for mounting purposes (always cut out your largest pieces first). Standard #4-40 machine screws/nuts were used to fasten the aluminum angles together.

The outward-facing legs at the top of the frame holding the linear actuators were flush (no separation distance) with the legs of the angles of the outer frame. Small end pieces of aluminum angles were used to mount the heat gun to the interior support angles that ran from the rear of the frame to the front. A power strip was mounted on the horizontal angle at the rear of the box. The residual aluminum angle material was fastened to the back of the frame to secure a switching power supply recycled from an unwanted/outdated computer.

## 2.2 Components and actuators

### 2.2.1 Spin coater

The spin coater was made from a brushless direct-current (BLDC) motor of a hard drive taken from a retired desktop computer. Any BLDC motor with the ability to reach high revolutions-per-minute (RPM) can be used for the rotational actuator of the spin coater, but 1) the hard drive frame is large and can be made into a cart by attaching wheels and 2) there are previously documented procedures for successfully recycling the BLDC motors from computer hard drives for use as spin coaters. [19–22] The casing was removed followed by the read/write arm components and platter. Depending on the institution, protocols can be in place that make it difficult to dispose of outdated university property, where many units may be in queue for removal over an extended period of time. Thus, it can be easy for educators/students with limited budgets to take advantage of this local stockpile of outdated materials/devices held by a university's Information Technology Services.

A 4.5" diameter, miniature Bundt pan was used as the solvent catch bowl. A 1 1/4" diameter hole was drilled through the center of the Bundt pan from the rear using a hole-saw drill bit. The drilled edge of the Bundt pan was filed smooth and mounted on the hard drive frame with the motor viewed through the center hole. A hollow cylindrical spacer was fixed over the rotor to raise the spin coater plate above the inner lip of the catch bowl. A circular section on the hard drive casing was placed over the spacer for use as a spin coater plate and fastened to the rotor with three screws. Various models of magnetic hard drives are in common use, where this particular model attached the disk with a cap using three off-axis screws as opposed to a single centered screw.

Some designs for hard-drive spin coaters require a small hole to be drilled through the rotational axis of the rotor; these designs can use an external vacuum pump to create a seal with the sample substrate. [19] Pumps can be costly and not all recycled hard drives can support this capability. Therefore, we used three $\sim 0.5\,\mathrm{cm}$ pieces of Scotch brand mounting tape instead of a vacuum seal. When a rotor speed above $6000\,\mathrm{rev/min}$ was required, we replaced one of the pieces of Scotch brand mounting tape with a piece of Gorilla Glue Heavy Duty mounting tape. Note that glass substrates were easily removed from the spin coater plate when only Scotch brand mounting tape was used or when only one piece of the heavy duty mounting tape was used, but glass substrates would often break from the stress applied to remove them when all three pieces of mounting tape were heavy duty.

A line of lead-based solder was ironed onto the bottom of the spin coater plate's outer edge to form a lip. The dense lead-based lip served two purposes; 1) residual solvent after spinning dripped from the lip into the solvent catch bowl instead of flowing into the motor, and 2) the heavy lip increased the plate's moment-of-inertia to reduce angular jerk when accelerating/decelerating.

### 2.2.2 Solvent deposition

Spin coating layers of polymeric materials requires the polymers to first be dissolved in solvents. The polymers in spin coated multilayers are limited to pairs soluble in orthogonal solvents, [18] where an orthogonal solvent only dissolves only one of the two polymers used during processing. During the spin cycle, the radial acceleration of the rigid substrate allows the viscous flow of material to coat the flat surface. [23] The coated film can be made with relatively uniform thickness across much of the substrate over a broad range of spin parameters. [24]

In addition to multilayer reflectors with quarter-wavelength layer thicknesses, constructive interference in the reflected light can also occur at greater layer thickness intervals. [25] For multilayers composed of high/low refractive index bilayers, the optimal thickness $t$ for a layer to reflect light at a wavelength $\lambda$ follows as $t = (m + 1/2)\,\lambda/2n$, where $n$ is the refractive index of a single layer and $m = 0, 1, 2, \ldots$. A multilayer Bragg grating typically consists of tens to hundreds of individual layers.

The polymer/solvent solutions were deposited on the substrate by actuated syringes above the spin plate. The syringes were mounted using spare aluminum angle material and 3/4" split ring hangers as shown in Figure 1(a). The syringes were surrounded by pipe insulation foam before being clamped tightly into position by the split ring hangers. The heavy duty linear actuators were the most expensive components used in the automated system. Optical mounts were connected to the actuator and two washers were tightened around the syringe plunger to hold it in place. Other designs for multilayer spin coaters employ dispensers fed through tubes. [26] Our system was designed to use 12 ml disposable syringes which reduces the amount of cleaning required after operation; however, the chosen length of the linear actuators (20 cm travel length) allows for much larger syringes, if required.

### 2.2.3 Annealing

In addition to macroscopic mechanical parameters such as angular speed, duration of radial acceleration, and higher-order time derivatives of the rigid substrate's angular position (*e.g.*, jerk), open-system interactions of the polymer/solvent solution with the local environment are also of great importance. The solvent in the polymer/solvent solution must be allowed to evaporate prior to subsequent layers being deposited. Therefore, adequate time during the spin process, annealing time, and cooling time must be allowed for the solvent in each layer to evaporate. Ambient conditions such as temperature and relative humidity can directly affect the time required to evaporate the solvent in each layer. [27, 28]

Thermal annealing can significantly raise the temperature of the sample and surrounding environment. Therefore, the annealing steps were performed away from the dispensing needles by moving the spin coater to the rear of the device. To make the spin coater assembly mobile, sets of wheels were fastened to the front and rear of the spin coater which rolled along the legs of aluminum angles at the base of the apparatus. A belt-fed linear actuator driven by a DC motor, recycled from a Canon MG2500 series printer, was attached to the hard disk's frame and oriented to move towards and away from the front of the multilayer spin coater system. The use of a belt-driven linear actuator was due to availability only, where most linear actuators found in computer parts (*e.g.*, optical disk tray or laser positioner from a CD/DVD drive) would work well for the task of translating the spin coater assembly between the dispensing and annealing positions. In the back-most position, a 300 W heat gun was directed straight down and fixed to aluminum angle supports. The placement of the aluminum mounting angles used to hang the heat gun are shown in Figure 1(b). The heat gun was mounted in the optimal position and orientation to heat the sample based on the distance that the spin coater was able to move away from the deposition position. A 12 V cooling fan was affixed to vertical rail and oriented towards the annealing position, where the fan was used to cool the sample following thermal annealing.

### 2.2.4 Controls and automation

An Arduino with a ZYLTech CNC shield was loaded with the GRBL library and fastened to an interior aluminum angle. The $X$ and $Y$ axes controllers on the CNC shield were connected to the stepper motors that drove the linear actuators used for depositing the polymer/solvent solutions on the substrate. A second Arduino was loaded with the code given in Section S2 of the supplementary sections. The pulse width modulation (PWM), pin 5, acted as a virtual DC signal to an electronic speed controller (ESC) which controlled the spin coater's rotor. The ESC did not require a motor with Hall sensors, but instead, it determined the rotor speed by sensing the induced *emf* of the stator winding as the magnetic flux changed during rotation of the motor's permanent magnets. This type of ESC is colloquially referred to as "sensorless." The remaining logic pins were connected to 3.3 V relays with optocouplers. The relays connected to pin 2 controlled the heat gun by closing a circuit with a 120 V AC source, and the relay connected to pin 8 closed a circuit with a 12 V source in series with a computer cooling fan. The relays connected to pins 4 and 7 controlled the tray movement by closing a ±12 V sourced circuit with the DC motor. Both Arduinos were connected to the USB ports of a Raspberry Pi 3B with a 5" touch screen which was fastened to the apparatus frame.

## 2.3 Polymers and solvents

Low-cost, commercially available polymers and solvents were chosen for this study. Polystyrene (PS) was recycled from an expanded polystyrene cooler used to ship frozen goods. Klean Strip® toluene was used to dissolve the PS. Cellulose acetate (CA) was taken from grafix™ acetate overlay sheets.

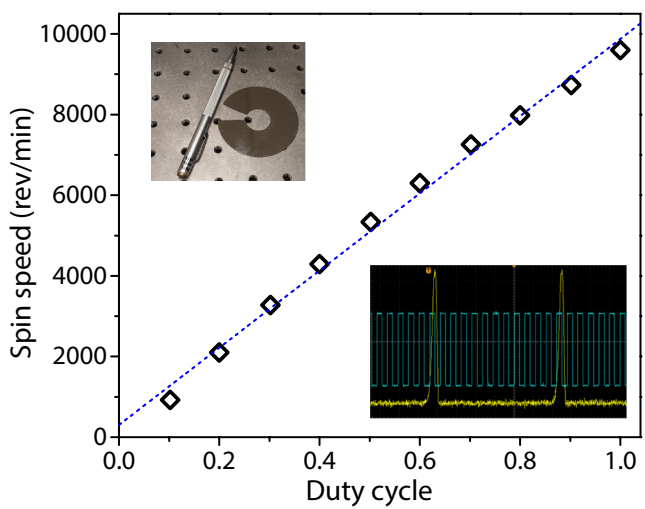

Figure 2: (Color online) The rotor speed as a function of duty cycle. The upper-left inset shows a hard disk platter with a section removed with a glass cutter tool. The small piece was attached to the spin coater plate in order to determine the rotor speed by reflecting a CW laser beam. The lower-right inset shows an oscilloscope screenshot of showing the tachometer signal (blue line) and the photodetector signal from the laser beam reflected from the spin coater plate (yellow line).

Diacetone alcohol (DAA) purchased from Consolidated Chemical & Solvents, LLC was used to dissolve the CA.

## 3 Characterization

The relative motion "G91" GRBL command was used to move the linear actuators position as shown in Section S3 of the supplementary sections. The linear interpolation code "G01" with a feed rate of "F070" was determined to be the optimal command sequence for 16 gauge dispensing needles, where the fluid speed exiting the tip was low enough to avoid splashing while maintaining an acceptable volume flow rate. The feed rate should be adjusted if a different diameter syringe and/or dispensing needle is chosen. Because the needle can drip fluid due to the slow relaxation time from pressure built up after the linear actuator has reached its final position, an additional backwards actuation was introduced. Using the default units for the GRBL commands, we determined that a travel distance of 0.7 provided enough fluid to cover a 1"×1" glass slide without flowing over the substrate. After a few seconds, a backwards travel distance of 0.2 removed all dripping while keeping fluid in the dispensing needle. These parameters worked well for both polymer solutions. Note that a change in units, stepper resolution, thread pitch, viscosity, syringe size, dispensing needle gauge, and type of fluid will affect the numerical value for the travel distance commands. Thus, these parameters need to be established for individual prototype systems.

The angular speed of the spin coater depended on the PWM input to the ESC, where the duty cycle of the PWM acts as a virtual DC signal. The frequency of the PWM signal is ideally much greater than the step frequency signal to the BLDC from the ESC. Section S3 of the supplementary sections illustrates how to change the PWM frequency via a change in the clock speed; a good resource for changing the frequency of PWM pins on an Arduino can be found in Ref. [29]. The angular speed as a function of the input PWM duty cycle was determined with a digital oscilloscope using optical tachometry. Because data sheets may not be available for all BLDC motors and taking some apart can result in irreversible damage, we also cut a section of the hard drive platter to act as small mirror on the spin coater. By reflecting a collimated laser diode light beam off the surface of the mirrored disk every time it passed across the beam path, we were able to determine the time interval of a revolution *via* photodiode detector. The measured rotor speed as a function of the duty cycle is given in Figure

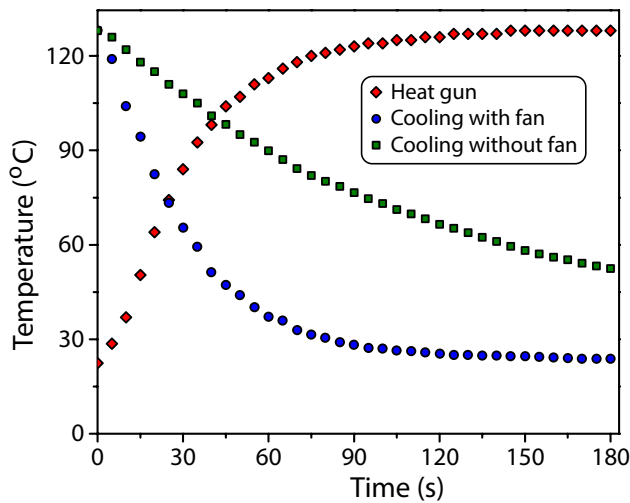

Figure 3: (Color online) The temperature measured by a type-K thermocouple as a function of time during heating and cooling periods of the annealing step.

2. The lower right inset shows a screen shot of the oscilloscope between the channel detecting the ESC tachometer and the channel detecting the photodiode. The tachometer revealed twelve cycles per one full rotation of the rotor. The maximum angular speed of the rotor for the hard disk connected to the spin plate is $\sim 9600\,\text{rev/min}$. The Arduino has 256 bit resolution (0-255) for the duty cycle which means that the angular speed can be selected with a resolution of $\sim 37.6\,\text{rev/min}$.

An option to anneal after each fluid deposition and spin cycle was included in the system. A 300 W, single-setting, heat gun was aimed downward at the sample (when positioned away from the deposition arms) with the heat gun's nozzle located $\sim 6\,\text{cm}$ above the sample surface. Note that a multi-setting heat gun could be used for increased functionality. The characterize and test the settings, the temperature at the sample location as a function of time was recorded with a type-K thermocouple after the relay was tripped to deliver power to the heat gun and is shown in Figure 3.

Cooling of the sample, substrate, and surrounding material can be a slow process. Because an automated system is in part designed to reduce the total time required to fabricate the multilayer materials, a cooling fan was introduced to further reduce the time period of the designated annealing steps. Figure 3 also shows the temperature as a function of time when cooled with and without the cooling fan.

Single layers of polymer films were spincoated from solution onto substrates and annealed. The samples were then shipped from Hawaii Pacific University to Youngstown State University. The samples were then scored and the thickness was measured with a stylus profilometer. The thicknesses measured with a profilometer as a function of rotor speed for different polymer/solvent concentrations for both CA and PS are shown in Figure 4. A simple model of the thickness $d$ versus the spin speed $\omega$ is fit to the data shown in Figure 4. The model follows the expected relationship $d \propto 1/\sqrt{\omega}$. [30, 31] The model fits the data well for low concentrations while the highest concentrations, $23\,\text{g/L}$ for PS and $29\,\text{g/L}$ for CA, appear to have the greatest residuals over the range of spin speeds.

## 4 Applications

Two applications for the automated multilayer spin coater system are presented in this section. The first describes a dielectric mirror and the second involves the fabrication of a DFB laser. Both applications use the same polymer/solvent pairs.

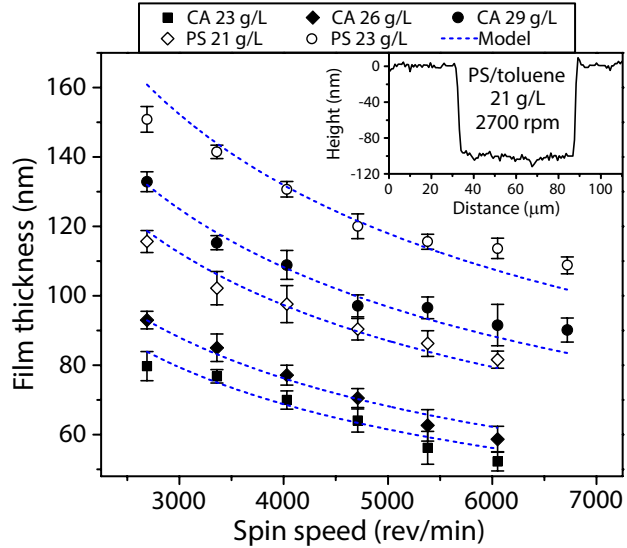

Figure 4: (Color online) Layer thickness as a function of spin speed for differing PS/toluene and CA/DAA concentrations. The inset shows an example stylus profilometer measurement of the layer thickness vs. horizontal position from which the data in the main graph was derived.

## 4.1 Polystyrene/cellulose acetate Bragg mirrors

A square glass slide with 1" side lengths was secured to the spin coater in the same manner described in the previous section. Two 12 ml disposable syringes were secured below the dispensing arms. As determined from thin characterization in the previous section, the first syringe contained 23 g/L PS/toluene and the second syringe contained 29 g/L CA/DAA. The linear actuator arms were repositioned and the syringe plungers were clamped between two washers. A paper towel was placed beneath the dispensing needles and the actuators were moved downwards until fluid spurted from both needles. A short reversal of the actuators (X-0.2 and Y-0.2) followed so that any slow drip from the dispensing needles was eliminated. The spin coater was repositioned with the substrate directly underneath the dispensing needles.

The Python script shown in Section S3 of the supplementary sections was executed until 15 PS/CA bilayers (30 individual layers) were deposited. The processing parameters for Bragg reflector fabrication follow:

- PS/toluene spin speed: **7400 rpm**

- CA/DAA spin speed: **3900 rpm**

- Spin duration for both solutions: **45 s**

- Annealing duration for both solutions: **120 s**

- Fan-cooling duration for both solutions: **150 s**

After all layers were deposited, the glass substrate holding the film was removed from the spin plate by slowly prying it away with a small tool. Any large mounting tape pieces still stuck to the bottom of the glass substrate were mechanically removed. The bottom of the substrate was then wiped clean of any residual adhesive with acetone and methanol.

Transmission spectra were recorded with an Ocean Optics 4000 USB spectrometer. The white-light probe had a beam diameter of $\sim 2$ mm. A glass substrate without any coating was placed in the gap and used as a reference. The transmission spectrum of the multilayer film (still adhered to the glass substrate) is shown in Figure 5. The transmission spectrum shown in Figure 5 is well-formed, but the reflection band became unesthetic when probed near the edge of the film as shown in Section S4 of the supplementary sections.

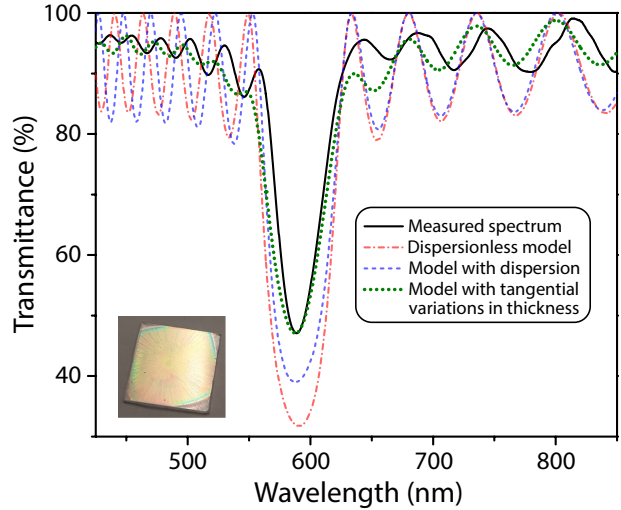

Figure 5: (Color online) The transmittance as a function of wavelength for a 15 bilayer PS/CA multilayer film. The transmittance is compared to three transfer-matrix method models with increasing complexity.

The transfer-matrix method [32–34] was used to model the transmittance of the multilayer film. The transmittance determined with the spectrometer was compared to three separate models. The first model assumed that there were no variations in the thickness of each type of material. It also assumed that the refractive index of both materials was constant. This simple model greatly overestimated the reflectance in the band region and did not accurately describe the transmittance oscillations at shorter wavelengths.

The second model included dispersion in the refractive index of both materials, but it still assumed that the layer thicknesses were perfect throughout the entire multilayer reflector. The refractive indices as functions of free-space wavelength were estimated with a Cauchy model, $n \approx A + B/\lambda^2$, where $A$ and $B$ are fit parameters and $\lambda$ is the wavelength. The CA refractive index was determined by fitting the Cauchy model to data in Ref. [35]. Similarly, the PS refractive index was determined using data from Ref. [36]. The Cauchy model parameters for CA were determined to be $A = 1.46$ and $B = 5.0 \times 10^3 \,\text{nm}^{-2}$. Likewise, the parameters for PS were estimated to be $A = 1.56$ and $B = 1.1 \times 10^4 \,\text{nm}^{-2}$ over the visible spectrum. The model that included dispersion overestimated the reflectance in the band region, but to a lesser extent than the dispersionless model. The amplitude of transmittance oscillations outside the reflection band was still greatly exaggerated in the model. The frequency of oscillations, however, increased at shorter wavelengths when compared with the dispersionless model. In comparison to the dispersionless model, the second model was able to describe the observed transmission spectrum to a better degree, but it was still far from accurate.

A third model included dispersion and also assumed that layer thicknesses fluctuate tangentially across the film. Because the probe beam had a relatively large spot size on the sample, variations in layer thicknesses over the beam spot can affect the measured transmittance. Thus, the third model assumes random thickness variations about the nominal layer thickness of each material. To simulate tangential variations, the model assumes many cross-sectional areas in the film with different random variations, where the spectra of each small patch of area are summed together and divided by the total number of sections. We chose to calculate and average 100 subsections of area being illuminated. The random variations were assumed to be Gaussian about the mean for each material. The model shown in Figure 5 assumes a mean layer of 96 nm for both CA and PS and an 8.5 % standard deviation in layer thicknesses. The third model, which assumes that the layer thicknesses fluctuate tangentially across the multilayer film, provides a good fit to the observed transmission spectrum.

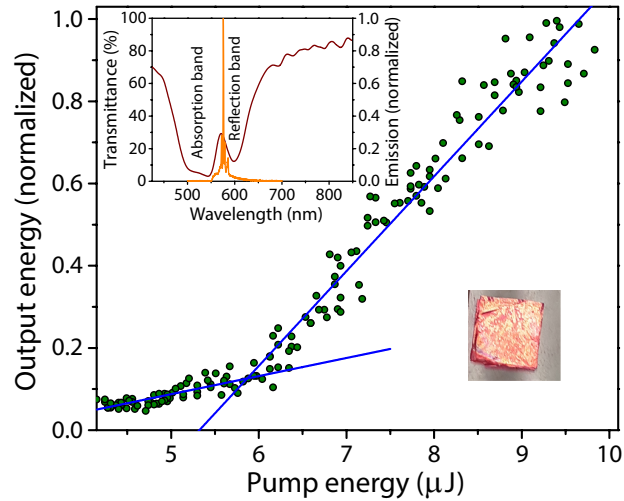

Figure 6: (Color online) The output power as a function of pump power for the PS/R6g-CA multilayer DFB laser. The upper left inset shows the laser transmission and emission spectra. The lower right inset is an image of the multilayer DFB laser.

### 4.2 Distributed feedback laser

As a demonstration of the spin coater's layer reproducibility, a multilayer DFB laser was fabricated using the same Bragg reflector recipe described above. For the DFB laser, the CA/DAA fluid was doped with rhodamine 6g (R6g) at a concentration of 1 wt.% with respect to the CA polymer. For 29 g/L of CA, the concentration of R6g with respect to the DAA volume was 290 mg/L. The Bragg bilayers have a refractive index difference $\Delta n_{PS/CA} \sim 0.1$, where a 30-layer film is typically insufficient for DFB lasing. For example, DFB lasers fabricated from the terpolymer tetrafluoroethylene-hexafluoropropylene-vinylidene fluoride (THV) and the copolymer styrene-acrylonitrile (SAN) have been reported using 128 layers with a greater refractive index difference, $\Delta n_{THV/SAN} \sim 0.2$. [37] The 12 ml syringes used in this study do not hold enough fluid to spin such large numbers of layers in a single run. Rather than spin many layers that require multiple refilled syringes, a single run was performed to create a post-processable 30-layer film. The multilayer film was then cut with a razor blade into four equal-sized rectangles that were stacked on each other to increase the number of layers to 120. Each rectangle was glued together with an ultrathin adhesive layer which resulted in a 120-layer PS/R6g-CA DFB laser.

A 532 nm, 6 ns pulsed, Nd:YAG laser operating at 10 Hz was used to pump the multilayer DFB laser. The pump was reflected downward from a dichroic mirror and focused with a lens onto the multilayer DFB laser. The light emitted from the surface of the multilayer DFB laser was collimated by the lens and transmitted through the dichroic mirror. The emitted light passed through a 550 nm high-pass filter and onto the end of an optical fiber that fed into a spectrometer. The pump intensity was adjusted with a variable attenuator. The output power of the multilayer DFB laser as a function of pump power is shown in Figure 6. The plot is commonly referred to as a "J-curve" due to the shape caused by a laser's threshold behavior. The emission spectrum is shown in the inset of Figure 6 and overlayed with the transmission spectrum of the DFB laser. The absorption band of R6g is located next to the reflection band of the multilayer film. The laser emission wavelength is located at the highly dispersive, short wavelength, reflection band edge. The J-curve is in good agreement with previous plastic distributed Bragg reflector laser measurements at high pump intensities [14] and ASE measurements at low pump intensities. [40]

### 4.3 Instructional use

The system described in this paper can be used to create prototype Bragg reflectors for student-centered research projects and instructional laboratory courses. At Hawaii Pacific University, the modern physics laboratory course includes a coherent light assignment in which students fabricate a distributed feedback laser.

A neat dye-doped polymer should be fabricated with a single spin coated layer when initially investigating incoherent light emission. Note that properties such as emission spectrum and far-field interference can be contrasted between the incoherent and coherent light in the written lab report. Using a low-cost, high-quantum-yield dye such as R6g in cellulose acetate is suitable for a second-harmonic Nd:YAG pump. Spin coating a thick layer onto a glass slide and focusing the pump beam onto a line with a cylindrical lens is the best method for generating amplified spontaneous emission (ASE). Because the ASE threshold depends on the length of spontaneously emitted photons traveling through the gain medium (the part of the dye-doped polymer being illuminated, a gain region should be broad. Using a cylindrical lens to focus a laser line onto the dye-doped polymer forms the gain region into a long cylinder-like shape, which i) allows for a clearly defined propagation direction and ii) efficiently utilizes most of the pump intensity available for the purposes of gain. Note that the neatness of the polymer film is important because impurities and roughness can sometimes result in random laser lines protruding from the ASE spectral profile. Fluorescence spectra can also be obtained from the same sample at very low intensities or the sample can be placed in the path of a beam with a circular cross section.

The second sample is the multilayer DFB laser which can be fabricated by the students. For a laboratory course meeting once a week, we recommend scheduling the laboratory assignment after the second week in the semester to allow for preparation. The solvent/polymer solutions (including any dye) should be made two weeks in advance. Although PS dissolves quite rapidly in toluene, the CA can be stubborn when trying to dissolve in DAA, where many cycles of heat and sonication can be required. Thus, the instructor may find themselves finishing up the process for CA/DAA outside of laboratory hours so that students may focus on the main laboratory assignment for that week. The DFB laser should be fabricated a week before the assigned laboratory. Because of the simple turnkey design, students can take a moment at the beginning of their regularly scheduled laboratory assignment to load the solvent/polymer solutions into the system and start the process. A 15 bilayer film should be finished before the end of a three-hour laboratory for inspection. Also note that the instructor is advised to run a prototype days before to ensure that the system's settings will result in a suitable DFB laser using the solvent/polymer solutions mixed in the previous week. The film can be lifted, cut, and stacked to create a DFB laser with many layers during the same preparation week which allows for transient mechanical behavior to dampen prior to beginning the coherent light laboratory assignment.

The third sample is used to illustrate laser emission from weak photon localization, which is commonly referred to as random lasing. By the time the students have taken fluorescence spectra, ASE spectra, and DFB emission spectra/J-curve/far-field diffraction image, there is little time remaining for a major undertaking like testing random lasing as a function of microparticle size or concentration. Thus, a simple proof-of-concept approach is recommended, where the easiest and most cost-effective method to show random lasing is by staining a rough surface and using high intensity pump pulses. Large samples of bovine bones can be procured from T-bone and porterhouse cuts of meat available at local grocery stores; bovine femur bones from marrow cuts can often be purchased from specialty meat suppliers. More information on random lasing from dyed bones are given in Refs. [38] and [39].

# 5 Discussion

The system has features that allow for a diverse set of recipes using various materials; however, some improvements are planned that could greatly improve the quality of the resultant multilayer films. Wheels were mounted to the hard-drive housing as a simple and cost effective means of moving the sample away from the dispensing syringes (often filled with volatile chemicals) for the annealing step. Mounting the hard-drive housing to a set of linear guide rails offers increased stability of the hard disk during the spin process. This design feature could result in a greater tolerance when mounting the spin plate for newly constructed systems by constraining wobble caused by an off-center spin plate. The rotor used in our system required three screws to mount the hard-drive platter which was later replaced by an elevated spin plate. Swapping the hard drive for a center-tapped rotor design would reduce the precision required to mount the spin plate, but the wobble would be compensated by the linear guide rail. Center-tapped rotors have been shown to hold substrates in place via vacuum systems, and therefore, adding a linear guide rail could both reduce wobble and remove the necessity for substrate adhesives.

As mentioned in the above paragraph, some costs were reduced by using inexpensive materials that could be replaced with more expensive alternatives. Likewise, more parts of the presented system could be replaced by low-cost materials with some sacrifice to performance. For example, the CNC arms used to push the syringe plungers were the most expensive items in the cost analysis shown in Section S5 of the supplementary sections. Replacing the CNC arms with possible alternatives such as recycled linear actuators used for positioning the lasers in Blu-ray/DVD/CD drives could be a viable alternative, especially when considering that several other parts of the system were recycled from desktop computers. We did not attempt this approach because the pitch on those types of linear actuators is large and the motors are relatively weak; however, either swapping out the smaller National Electrical Manufacturers Association (NEMA) motor for a larger NEMA motor or using a rail with a finer pitched threading could reduce the overall cost of materials.

Even with the above improvements to the mechanical system, poor sample quality can occur when solutions are not well-prepared. For the materials used in this paper, the PS/toluene solutions were the simplest to prepare at the concentrations shown in Figure 4. The CA/DAA was much more difficult to dissolve and required many cycles of heating, mixing, and bath sonication. Despite appearing fully dissolved, some marks were formed on the sample during the spin coating process. These marks could have been caused by microscopic debris in the atmosphere landing on the film during the spin coating process. The markings could have also been formed from poor dissolution of the polymer. Pushing the solution through a polytetrafluoroethylene (PTFE) filter prior to filling the dispensing syringes could potentially reduce the discontinuous markings in the film.

There are many options for screens suitable for Raspberry Pi models. The 5" touch screen used in our system was large enough to see the graphical user interface (GUI) of the operating system, but it was too small to easily use the keyboard interface to edit the Python script while viewing the changes to it. Thus, a keyboard and mouse were used despite the touch screen functionality. Using a larger touch screen would reduce the need for a keyboard and mouse. Alternatively, a screen that only requires an HDMI port would free many Raspberry Pi pins. Then many actuators can be controlled directly from the Raspberry Pi, where the Arduino Uno used to control the relays and BLDC driver's PWM could be removed from the system.

As a final note, there are no interlocks on the low-cost spin coater/annealer system. Thus, there should be precautions taken before running the system; mainly, the system should have intermittent supervision (it should not be run overnight without staff present). The relays should all have interlocks in the event that one does not trip to the proper setting. Overlapping interlocks should also be sent such as a temperature sensor. An interlock that senses final dispenser plunger position should also be employed so that a use does not try to move the CNC arm below the emptied plunger position. Lastly, the system should not be run while under a fume hood or air recycling unit that is housed in a shared facility; the annealing feature can add heat to the surrounding environment that could potentially ignite fumes from stored chemicals.

# 6 Conclusion

The spin coater and thermal annealer system was built to reproducibly fabricate multilayer photonic crystal prototypes at a relatively low cost. The system was built in a short time and the system components can be characterized using inexpensive sensors and equipment commonly found at an academic institution. The consumer grade polymers and solvents used to create the Bragg gratings were also inexpensive. The system required only minimal preliminary steps for the syringes (filling, inserting, and priming) and glass substrates (cutting, cleaning, and adhering) before starting the automated process. The system code given in Section S3 of the supplementary sections offers a degree of flexibility when creating new recipes and it only requires slight modifications to further the functionality, *e.g.*, Python command order can be changed to dispense while plate is spinning.

# Supplement Section S1: Additional images

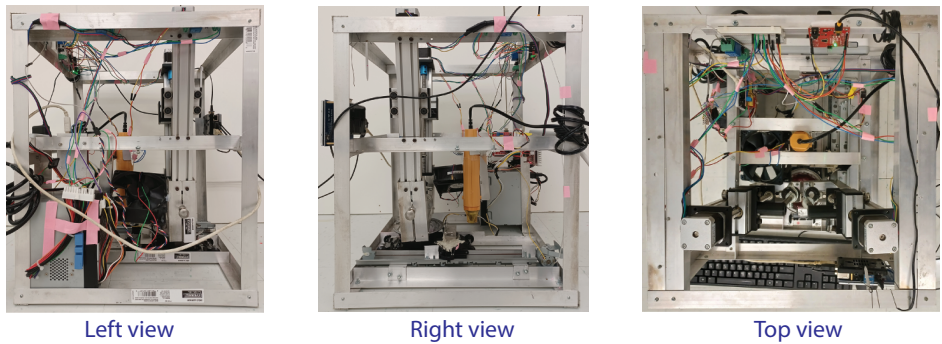

Figure 7: Pictures of the multilayer spin coater and thermal annealer from perspectives not shown in the published article.

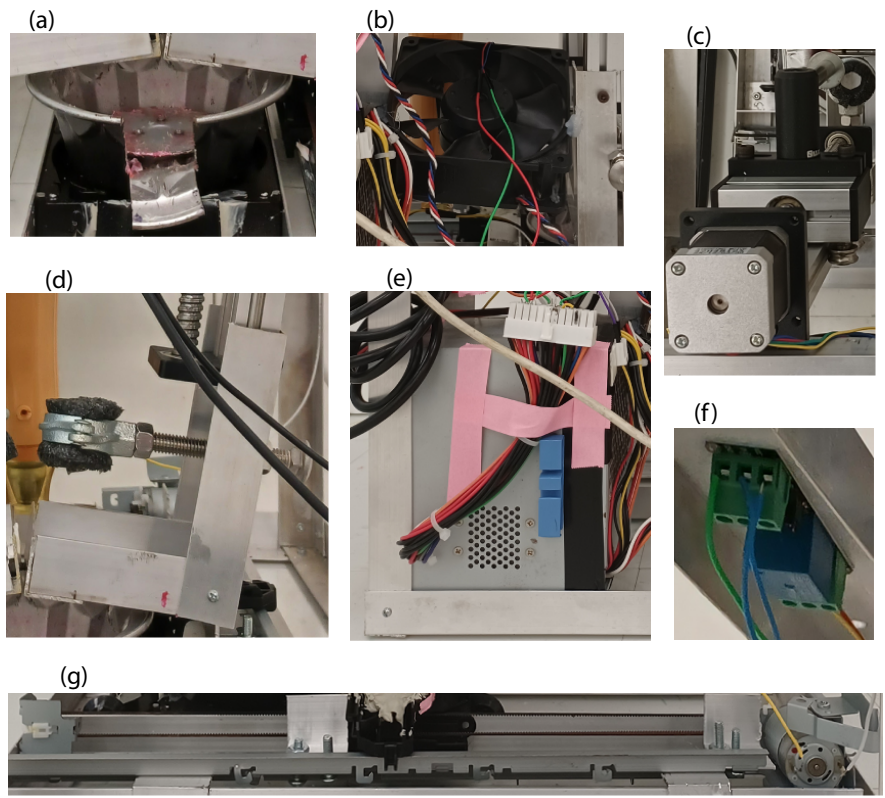

Figure 8: Pictures of some device components and assemblies. (a) The solvent catch bowl made from a Bundt pan has a notch cut out of the lip to allow the dispensing needles to pass through when the spin coater is translated between the dispensing and annealing positions. The spin coater plate with three screws used to mount it onto the BLDC motor. (b) The cooling fan recycled from a desktop computer. (c) The NEMA motor driving the linear actuator for dispensing fluid. (d) The syringe holder made from residual aluminum angles and a split ring hanger clasping pipe insulation. (e) The power supply recycle from a desktop computer. (f) A logic relay controlled by an Arduino. (g) A belt-fed linear actuator from a retired printer for translating the spin coater assembly.

## Supplement Section S2: Arduino code

The code uploaded to the Arduino microcontroller that ran the 1) spin coater, 2) platter movement, and 3) the heat gun is as follows:

```
int data;
int dataAold;
int dataBold;
int dataCold;
int dataDold;
int dataEold;
int dataA;
int dataB;
int dataC;
int dataD;
int dataE;
int pulwidmodA=5;
int relayB=2;
int relayC=4;
int relayD=7;
int relayE=8;
String dataget;
void setup() {
  Serial.begin(9600);
  TCCR0B = TCCR0B & B11111000 | B00000010;
  pinMode(relayB, OUTPUT);
  pinMode(relayC, OUTPUT);
  pinMode(relayD, OUTPUT);
  pinMode(relayE, OUTPUT);
  digitalWrite(relayB, LOW);
  digitalWrite(relayC, LOW);
  digitalWrite(relayD, LOW);
  digitalWrite(relayE, LOW);
  data = 0;
  analogWrite(pulwidmodA, data);
  dataAold = data;
  dataBold = 1;
  dataCold = 1;
  dataDold = 1;
  dataEold = 1;
}

void loop() {
  while (Serial.available())
  {
    dataget = Serial.readString();
  }
  char partid = dataget.charAt(0);
  dataget.remove(0,1);
  if (partid == 'A') {
    dataA = dataget.toInt(); }
  if (partid == 'B') {
    dataB = dataget.toInt(); }
  if (partid == 'C') {
    dataC = dataget.toInt(); }
  if (partid == 'D') {
    dataD = dataget.toInt(); }
  if (partid == 'E') {
    dataE = dataget.toInt(); }
```

```
55    if (dataA != dataAold) {
56      analogWrite(pulwidmodA, dataA);
57      dataAold = dataA;
58    }
59    if (dataB != dataBold) {
60      if (dataB == 1) {
61        digitalWrite(relayB, LOW); }
62      if (dataB == 2) {
63        digitalWrite(relayB, HIGH); }
64      dataBold = dataB;
65    }
66    if (dataC != dataCold) {
67      if (dataC == 1) {
68        digitalWrite(relayC, LOW); }
69      if (dataC == 2) {
70        digitalWrite(relayC, HIGH); }
71      dataCold = dataC;
72    }
73    if (dataD != dataDold) {
74      if (dataD == 1) {
75        digitalWrite(relayD, LOW); }
76      if (dataD == 2) {
77        digitalWrite(relayD, HIGH); }
78      dataDold = dataD;
79    }
80    if (dataE != dataEold) {
81      if (dataE == 1) {
82        digitalWrite(relayE, LOW); }
83      if (dataE == 2) {
84        digitalWrite(relayE, HIGH); }
85      dataEold = dataE;
86    }
87 }
```

## Supplement Section S3: Python script

The python code ran from the Raspberry Pi 3 B+ follows:

```
1  import time
2  import serial
3
4  ## Parameters (You may adjust these parameters)
5  NumberOfBilayers = 15;
6  RPM1 = 7400; #spin speed (in rpm) of solvent 1
7  RPM2 = 3900; #spin speed (in rpm) of solvent 2
8  anneal1 = 'yes'; #anneal after spinning solvent 1 ('yes' or 'no')
9  anneal2 = 'yes'; #anneal after spinning solvent 2 ('yes' or 'no')
10 anneal_after = 'no'; #anneal after all bilayers have been spun
11 acctime1 = 45; #the time period in seconds for it to spin layer 1
12 acctime2 = 45; #the time period in seconds for it to spin layer 2
13 heating_time1 = 120; #annealing time after spinning solvent 1
14 heating_time2 = 120; #annealing time after spinning solvent 2
15 heating_after = 0; #annealing time after all bilayers have been spun
16 cooling_time1 = 150; #time for sample to cool down after annealing solvent 1 layer
17 cooling_time2 = 150; #time for sample to cool down after annealing solvent 2 layer
18
19 ## Actuator settings (GRBL commands for dispensing)
20 cmdYforward = 'G91 G01 Y0.7 F070'
```

```python
21  cmdYback = 'G91 G01 Y-0.2 F070'
22  cmdXforward = 'G91 G01 X0.7 F070'
23  cmdXback = 'G91 G01 X-0.2 F070'
24
25  ## actuator command setup (Don't touch these)
26  cmdYforward = cmdYforward.strip('\n')
27  cmdYforward = cmdYforward.strip('\'')
28  cmdYback = cmdYback.strip('\n')
29  cmdYback = cmdYback.strip('\'')
30  cmdXforward = cmdXforward.strip('\n')
31  cmdXforward = cmdXforward.strip('\'')
32  cmdXback = cmdXback.strip('\n')
33  cmdXback = cmdXback.strip('\'')
34
35  ## PWM commands (Don't touch these)
36  dutycyc1 = int(255*RPM1/9600); #needs to be integer
37  strdutycyc1 = str(dutycyc1);
38  dutycyc2 = int(255*RPM2/9600);
39  strdutycyc2 = str(dutycyc2);
40  stringA = 'A';
41  pwmoff = 'A0'; #set to zero (not spinning)
42  pwmon1 = str(stringA + strdutycyc1);
43  pwmon2 = str(stringA + strdutycyc2);
44
45  ##Tray commands (Don't touch these)
46  relayCON = 'C2';
47  relayCOFF = 'C1';
48  relayDON = 'D2';
49  relayDOFF = 'D1';
50  tray_moving = 1.7; #time for spin coater to position
51
52  ##Heater/cooler commands (Don't touch these)
53  heatgunON = 'B2';
54  heatgunOFF = 'B1';
55  fanON = 'E2';
56  fanOFF = 'E1';
57
58  ##Open serial ports (adjust these for specific interface)
59  port = '/dev/ttyACM0'; #USB port connected to Arduino with CNC shield responsible
        for dispensing solvents via syringes
60  portB = '/dev/ttyUSB0'; #USB port connected to Arduino responsible for spinning
        and tray movement
61  baud = 115200;
62  baudB = 9600;
63  RedBoardSerial = serial.Serial(portB,baudB);
64  time.sleep(2);
65  CNCserial = serial.Serial(port,baud);
66  time.sleep(2);
67  #
68  ij = 1; #start counter for number of bilayers at one
69  while ij <= (NumberOfBilayers):
70      ## Begin 1st bilayer
71      # Dispense arm 1 liquid
72      CNCserial.write(cmdYforward.encode('ascii')+'\n'.encode('ascii'))
73      time.sleep(4)
74      CNCserial.write(cmdYback.encode('ascii')+'\n'.encode('ascii'))
75      time.sleep(3)
76      # Start PWM output
77      RedBoardSerial.write(pwmon1.encode());
```

```python
78      time.sleep(acctime1)
79      RedBoardSerial.write(pwmoff.encode());
80      time.sleep(10)
81      if anneal1 == 'yes':
82          ## Move tray under heater
83          RedBoardSerial.write(relayCON.encode());
84          time.sleep(tray_moving)
85          RedBoardSerial.write(relayCOFF.encode());
86          time.sleep(2)
87          ## Heating/cooling
88          RedBoardSerial.write(heatgunON.encode());
89          time.sleep(heating_time1)
90          RedBoardSerial.write(heatgunOFF.encode());
91          time.sleep(4)
92          RedBoardSerial.write(fanON.encode());
93          time.sleep(cooling_time1)
94          RedBoardSerial.write(fanOFF.encode());
95          time.sleep(8)
96          ## Move tray under spin coater
97          RedBoardSerial.write(relayDON.encode());
98          time.sleep(tray_moving)
99          RedBoardSerial.write(relayDOFF.encode());
100         time.sleep(5)
101
102     ## Begin 2nd bilayer
103     # Dispense arm 1 liquid
104     CNCserial.write(cmdXforward.encode('ascii')+'\n'.encode('ascii'))
105     time.sleep(4)
106     CNCserial.write(cmdXback.encode('ascii')+'\n'.encode('ascii'))
107     time.sleep(3)
108     # Start PWM output
109     RedBoardSerial.write(pwmon2.encode());
110     time.sleep(acctime2)
111     RedBoardSerial.write(pwmoff.encode());
112     time.sleep(10)
113     if anneal2 == 'yes':
114         ## Move tray under heater
115         RedBoardSerial.write(relayCON.encode());
116         time.sleep(tray_moving)
117         RedBoardSerial.write(relayCOFF.encode());
118         time.sleep(2)
119         ## Heating
120         RedBoardSerial.write(heatgunON.encode());
121         time.sleep(heating_time2)
122         RedBoardSerial.write(heatgunOFF.encode());
123         time.sleep(4)
124         RedBoardSerial.write(fanON.encode());
125         time.sleep(cooling_time2)
126         RedBoardSerial.write(fanOFF.encode());
127         time.sleep(8)
128         ## Move tray under spin coater
129         RedBoardSerial.write(relayDON.encode());
130         time.sleep(tray_moving)
131         RedBoardSerial.write(relayDOFF.encode());
132         time.sleep(5)
133
134     ## Increase counter
135     ij = ij+1; #increase counter by 1
136
```

```
137 ## Move tray away from residual solvent drip from syringes
138 RedBoardSerial.write(relayCON.encode());
139 time.sleep(tray_moving)
140 RedBoardSerial.write(relayCOFF.encode());
141 time.sleep(2)
142 ## Anneal step of multilayer stack
143 if anneal_after == 'yes':
144     ## Heating
145     RedBoardSerial.write(heatgunON.encode());
146     time.sleep(heating_after)
147     RedBoardSerial.write(heatgunOFF.encode());
148
149 ## Close connection to Arduino Unos
150 RedBoardSerial.close()
151 CNCserial.close()
```

## Supplement Section S4: Transmission spectrum profile of Bragg reflector

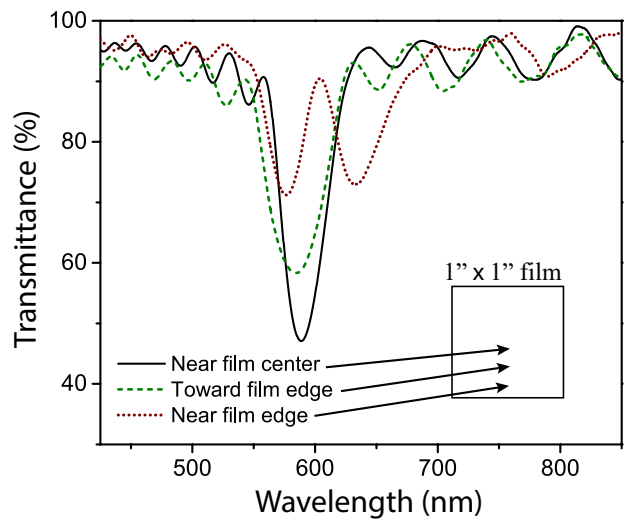

Figure 9: Transmission spectrum for three different locations on a spin coated, 15-bilayer, Bragg reflector.

# Supplement Section S5: Cost of Materials

Table 1: Condition and cost of materials in USD

| Item | number | Condition | Price per item | Cost |
|---:|:---:|:---:|:---:|:---:|
| 1"×96" Al angles (1/8") | 4 | New | $28 | $112 |
| Hard drive BLDC | 1 | Recycled | $0 | $0 |
| Arduino Uno (or equiv.) | 2 | New | $23 | $46 |
| ZylTech CNC Shield | 1 | New | $15 | $16 |
| Raspberry Pi 3B+ | 1 | New | $40 | $40 |
| 3.3 V relay | 4 | New | $3 | $12 |
| Printer belt actuator and DC motor | 1 | Recycled | $0 | $0 |
| 20 cm CNC linear actuator | 2 | New | $120 | $240 |
| Computer power supply | 1 | Recycled | $0 | $0 |
| Split ring hangers | 2 | New | $7 | $14 |
| 4.5" miniature Bundt pan | 1 | Repurposed | $0 | $0 |
| Power strip | 1 | New | $10 | $10 |
| Keyboard | 1 | Used | $0 | $0 |
| Mouse | 1 | Used | $0 | $0 |
| 5" Touch screen | 1 | New | $38 | $38 |
| Sensorless BLDC ESC | 1 | New | $20 | $20 |
| 300 W heat gun | 1 | New | $12 | $12 |
| 140 mm computer fan | 1 | Recycled | $0 | $0 |
| Various optical mounts | 6 | Repurposed | $0 | $0 |
| Wheels for hard drive | 4 | Recycled | $0 | $0 |
| **Total estimated cost** | | | | **$560** |

# Acknowledgments

The authors thank Pat Allen and the College of Natural and Computational Sciences staff for educational laboratories at Hawaii Pacific University for their support *via* supplies and testing equipment.